**Title:** Spectrum of genetic diversity and networks of clonal organisms

Alejandro F. Rozenfeld**\***, Sophie Arnaud-Haond**†**, Emilio Hernández-García**\***, Víctor M. Eguíluz**\***, Manuel A. Matías**\***, Ester Serrão**†** and Carlos M. Duarte**§**

**Author affiliations:**

**\*** Cross-Disciplinary Physics Department, IMEDEA (CSIC-UIB), Instituto Mediterráneo de Estudios Avanzados, Campus Universitat de les Illes Balears, 07122 Palma de Mallorca, Spain
**†** CCMAR, CIMAR-Laboratório Associado, Universidade do Algarve, Gambelas, 8005-139, Faro, Portugal
**§** Natural Resources Department, IMEDEA (CSIC-UIB), Instituto Mediterráneo de Estudios Avanzados, C/ Miquel Marques 21, 07190 Esporles, Mallorca, Spain

**Abstract**
Clonal reproduction characterizes a wide range of species including clonal plants in terrestrial and aquatic ecosystems, and clonal microbes, such as bacteria and parasitic protozoa, with a key role in human health and ecosystem processes. Clonal organisms present a particular challenge in population genetics because, in addition to the possible existence of replicates of the same genotype in a given sample, some of the hypotheses and concepts underlying classical population genetics models are irreconcilable with clonality. The genetic structure and diversity of clonal populations was examined using a combination of new tools to analyze microsatellite data in the marine angiosperm *Posidonia oceanica*. These tools were based on examination of the frequency distribution of the genetic distance among ramets, termed the spectrum of genetic diversity (GDS), and of networks built on the basis of pairwise genetic distances among genets. Clonal growth and outcrossing are apparently dominant processes, whereas selfing and somatic mutations appear to be marginal, and the contribution of immigration seems to play a small role in adding genetic diversity to populations. The properties and topology of networks based on genetic distances showed a "small-world" topology, characterized by a high degree of connectivity among nodes, and a substantial amount of substructure, revealing organization in sub-families of closely related individuals. The combination of GDS and network tools proposed here helped in dissecting the influence of various evolutionary processes in shaping the intra-population genetic structure of the clonal organism investigated; these therefore represent promising analytical tools in population genetics.

# 1. Introduction

The considerable progress achieved during the last decades in molecular biology and biotechnologies has greatly enhanced the potential of molecular markers for studying the process of evolution in natural populations in the framework of population genetics. As the empirical basis for population genetics is broadened, it is increasingly clear that the theoretical constructs under which population analyses are traditionally conducted involve assumptions that are often violated in natural scenarios, such as the random mating, equilibrium (Wright 1931), and non-overlapping generations commonly assumed for interpreting statistics on population genetic composition and structure (Hey & Machado 2003).

The summary statistics used in population genetics to estimate relevant parameters such as departure from panmixia or the population structure indeed rely on theoretical and mathematical models that involve the adoption of a somewhat narrow range of underlying parameters and demographic models; this greatly limits the scope of demographic situations that can be accurately explored (Hey & Machado 2003). Among others, three examples in which the assumptions underlying the classical Wright-Fisher model (Wright 1931) of population genetics are violated are the cases of endangered and of invasive species, as well as pathogenic species exhibiting recurrent fluctuations in population size linked to epidemiologic events. These are among the type of species most studied in evolutionary ecology or molecular epidemiology, precisely because they exhibit population dynamics that strongly depart from equilibrium, which very much limits the interpretation of classical population genetics statistics. These constraints on the application of conventional metrics of population genetic structure are even more evident for clonal organisms, the characteristics of which challenge the notions of effective population size and generation time (Orive 1993; Yonezawa et al. 2004). Moreover, a clear concept of the unit (genetic individual) on which evolutionary forces are acting is lacking (Fischer & Van Kleunen 2002; Orive 1995) for clonal species.

Examination of the genetic structure of populations is rooted in a comparison of the extent of genetic variability among individuals within populations, as well as differences among the populations, typically assessed using appropriate molecular markers applied to statistically-representative samples of individuals. Hypervariable markers such as microsatellites are commonly considered to be the markers of choice in assessing the genetic variability and structure of populations. This is particularly true in the case of clonal organisms, for which they allow, in a given sample, proper assessment of the individual level through the isolation of distinct multi-locus genotypes or lineages, which is a prerequisite for estimating variability and structure in clonal populations. Furthermore, most evaluations of genetic composition of clonal (and non-clonal) populations are based on summary statistics, such as heterozygosity or fixation indices, that do not consider the distribution of genetic distances among the sampled individuals. In fact, except in the case of co-ancestry coefficients mostly used to assess spatial autocorrelation, inter-individual distances are not commonly used in the literature (Douhovnikoff & Dodd 2003; Meirmans & Van Tienderen 2004).

The depiction of a population structure as a concerted representation of the genetic distances between agents indicates that a network approach is suitable for examination of the genetic structure of populations, in which the links between agents depend on the genetic difference between them. The use of networks to graphically represent genetic relationships has emerged as a useful tool in cases in which the nodes are haplotypes (Cassens et al. 2003; Excoffier & Smouse 1994; Morrison 2005; Posada & Crandall 2001; Rueness et al. 2003; Templeton 1992) or populations (Dyer & Nason 2004). Here we propose a network representation of the genetic structure of populations of a clonal organism, focused on the genetic individuals (in clonal plants designed as "genets", extending clonally by growing new shoots also called "ramets") as the interacting agents. We represent the intra-population genetic similarities among the genetic individuals as networks. In addition, we go one step further from the simple graphical representation of genetic relationships, and quantitatively analyze the properties of the resulting networks using tools (see Methods) successfully applied to other problems (Proulx et al. 2005), such as the characterization of food webs (Dunne et al. 2002a; Dunne et al. 2002b) and the analysis of protein (Jeong et al. 2001) or gene (Davidson et al. 2002) interactions.

We demonstrate this approach using a clonal seagrass species (*Posidonia oceanica*) for which a large data set, including microsatellite data for approximately 1500 shoots sampled from 37 populations across the Mediterranean, is available. We first propose a metric for genetic distances among individuals based on their observed multi-locus microsatellite genotypes, which allows us to describe the spectrum of genetic diversity within populations. We then explore the biological processes that yield the observed spectra, and use this knowledge to topologically represent the population as a network, from which structural diagnostics are then derived.

## 2. Available data
### 2.1. Model species

*Posidonia oceanica* is a clonal marine angiosperm restricted to the Mediterranean Sea, where it develops extensive meadows ranging from 0 to 40 m in depth (Hemminga & Duarte 2000). It is a very slow-growing organism, with the clones extending horizontally through the growth of rhizomes at approximately 2 cm year$^{-1}$, developing shoots (ramets, the individual module repeated to develop clones) at intervals of approximately 5–10 cm. This monecious species (i.e., both male and female flowers on the same shoot) (Hemminga & Duarte 2000) is characterized by sparse episodes of sexual reproduction. Individual *P. oceanica* shoots (ramets) live for up to 50 years, and the clones have been aged to over 1000 years (Hemminga & Duarte 2000). The plants are experiencing basin-wide decline and are subject to specific protection and conservation measures (Marbà et al. 1996; Marba et al. 2005; Moreno et al. 2001).

### 2.2. Multi-locus microsatellite genotypes

Approximately 40 *P. oceanica* shoots were sampled in each of 37 localities ranging, from west to east, from the Spanish Mediterranean Coast to Cyprus (Table 2, Supplementary information), encompassing a distance range of approximately 3500 km. In all meadows, shoots were collected at randomly drawn coordinates across an area of 20 m×80 m. Then a meristem portion of each shoot was removed, desiccated and preserved in silica crystals.

Genomic DNA was isolated following a standard CTAB extraction procedure (Doyle & Doyle 1987). The 37 meadows were analyzed with the most efficient combination (Alberto et al. 2003; Arnaud-Haond et al. 2005) of seven nuclear markers, using the conditions described by Arnaud-Haond et al. (Alberto et al. 2003; Arnaud-Haond et al. 2005). This set of microsatellite markers allows the unambiguous identification of clonal membership (Arnaud-Haond et al. 2005).

To avoid scoring errors, which would typically generate very small apparent genetic dissimilarities among individuals actually sharing the same multi-locus microsatellite genotype and would thus affect our estimates of genetic distance, all ramets with a distinct genotype for only two or one alleles were re-genotyped for those loci to ascertain their dissimilarity or to correct for genotyping errors. The clonal or genotype diversity of a meadow was estimated as:

$$R = \frac{G-1}{N-1}$$

where $G$ is the number of multilocus genotypes discriminated (considered as as many distinct genets) and $N$ is the number of samples (i.e. ramets) analyzed for the meadow.

The spatial autocorrelation in those meadows was tested for using the kinship coefficient proposed by Ritland (1996; 2000) and the $Sp$ statistics proposed by Vekemans and Hardy (2004) . Using the slope of the autocorrelogram ($\hat{b}_f$) and the average kinship in the smallest distance class ($\hat{F}_1$), the $Sp$ statistics is described as follows:

$$Sp = \frac{-\hat{b}_f}{1-\hat{F}_1}$$

The significance of $Sp$ values is tested for using a 1000 permutation test, assigning randomly one of the existing coordinates to each genet at each step (Arnaud-Haond & Belkhir in press).

## 3. Similarity metrics

In order to characterize the genetic structure of the different populations, some measure of genetic similarity among ramets is needed yielding a distance of 0 between identical genotypes. In our particular case, where the underlying data are based on multi-locus microsatellite genotypes, one ramet is characterized by a series of pairs of microsatellite repetitions at $k$ loci, with $k=7$ in our case.

More specifically, the genotype of a particular ramet, called $A$, is represented as:

$$A = (a_1, A_1)(a_2, A_2)\ldots(a_k, A_k),$$

where $a_i$ and $A_i$ are the allele length (in number of nucleotides) in both chromosomes at locus $i$.

Given a second ramet, $B$, with genotype

$$B = (b_1, B_1)(b_2, B_2)\ldots(b_k, B_k),$$

we define a dissimilarity degree between $A$ and $B$ at locus $i$ as:

$$d_i(A,B) = \min(|A_i - B_i| + |a_i - b_i|, |A_i - b_i| + |a_i - B_i|)$$

,

which provides a parsimonious (i.e., minimal) representation of the genetic distance, understood as the difference in allele length, between samples *A* and *B*. This distance is somehow similar to the Manhattan distance, defined in geometry as the distance between two points measured along axes at right angles, as if we order the alleles at each locus in such a way that $a_i<A_i$ and $b_i<B_i$, then the min function always select its first argument. We define *genetic distance* among ramets by averaging the contributions from all loci:

$$D(A,B) = \frac{1}{k}\sum_{i=1}^{k} d_i ,$$

which provides the degree of global dissimilarity between *A* and *B*. Since we have $D(A,A) = 0$, genetically identical individuals (clones, or genets) are at zero distance according to this definition.

To the best of our knowledge, the genetic distance metric *D* proposed here to characterize dissimilarities among diploid organisms has not been formally described as yet in the literature. It is, however (P. Meirmans, personal communication), the distance definition implemented for calculating distances among diploid organisms in the widely used genetic software GENOTYPE (Meirmans & Van Tienderen 2004).

### 4. Simulations

In this work we focused on the ranges of inter-individual genetic distances generated within the population to underline the intra-population factors (clonality, mutation, and mating system) that can account for such distances.

In order to estimate in particular the impact of the mating system we performed computer simulations to explore the genetic distance between parents and offspring. For each meadow, we started the simulation with the ramets sampled, and generated new sets of virtual ramets by implementing separately two classes of reproductive events: sexual reproduction within clones (i.e., selfing) or among genetically different parents (i.e., outcrossing).

In the simulations modelling outcrossing, we randomly picked pairs of parents from the considered meadow and generated by sexual reproduction 100 new individuals, constituting the first generation. Sexual reproduction consisted in the construction a new multi-locus genotype (offspring) by randomly selecting, for each locus, one of the two microsatellite repeats present in each of two distinct parental genets at that locus.

This procedure was repeated by selecting parents from the first generation to produce a second generation, then we picked parents from the second one to produce a third generation and so on up to 12 generations. Once a new generation has been produced, we computed the distribution of genetic distances between its individuals and their corresponding ancestors placed at the original population (generation 0). The resulting distributions were characterized by their mean and standard deviation, and the process repeated 100 times, to improve by averaging the determination of the mean intergeneration distances for each meadow.

The simulations modelling selfing were similar, except that a single parental genet was selected from the meadow to produce each offspring by random recombination of its two microsatellite repeats present at each locus. In this case only one generation of 100 offspring was produced, and the whole process was repeated 100 times to better estimate the mean selfing distance between parent and offspring.

### 5. Network analysis

In mathematical terms, a network is represented by a graph. A graph is a pair of sets G={P,E}, where P is a set of N nodes and E is a set of edges connecting the nodes. As explained below, we will analyze networks, associated to each population, in which the nodes are the genets, i.e., the different multilocus genotypes found at the meadow, and the links are established among genets at a genetic distance smaller than a threshold $D_{th}$. Each edge connects only two nodes ($P_i$ and $P_j$), and therefore can be assigned a weight or length equal to the distance or degree of dissimilarity between them $D(P_i,P_j)$. Depending on the maximum value of the distance ($D_{th}$) allowed between two nodes for them to be connected, the range of possible networks is between a fully connected network (when all distances are accepted, and therefore all individuals are connected), or a network in which only identical nodes are connected ($D_{th}=0$). Here we chose to study for each population the network built by using as a threshold the average distance, $\delta_{oc}$, found between parents and offspring in the simulations performed in that meadow to illustrate the pairwise genetic relationships within a "one generation" path. It is worth noticing here that the estimated percolation point of the networks is in most populations (33 upon 37), interestingly close to the $\delta_{oc}$ but slightly lower in general (regression equation is y=-0.36+ 1.03 x, $r^2$=0.78 data not shown). The percolation point in a network is defined as the point (in our case, the value of the genetic distance) at which the largest connected part of the network becomes fragmented, in a well defined mathematical sense (Havlin & Bunde 1996), i.e. most pairs of nodes are not connected by any possible link or path, and the network is therefore broken in several pieces made of small clusters or isolated nodes. This relationship between the percolation point and the estimated outcrossing distance suggests that precluding mechanisms generating distances slightly below the mean $\delta_{oc}$ would lead to the fragmentation of the entire system. In fact, the cases where the percolation point coincides with or is superior to the $\delta_{oc}$ estimated under the hypothesis of random rearrangement of gametes suggest the existence of departure from panmixia in the studied population. In some populations the networks indeed appear to be fragmented or partially fragmented into clusters (connected components), possibly illustrating the occurrence of sub-structure in the meadows analyzed. Inside a cluster, there is a path connecting any two nodes. On the contrary, there is no path connecting nodes belonging to different clusters. We define the quantity S as the size (number of nodes) of the biggest cluster in the network. We have S ≤ N.

### 5.1. Local properties.

The degree of connectivity $k_i$ of node $P_i$ is defined as the number of nodes linked to it (i.e., the number of neighbor nodes). If each of these neighbors were connected with all the others, there would be $E_i^{(\max)} = k_i(k_i-1)/2$ edges between them. The clustering coefficient $C_i$ of node $P_i$ is defined as:

$$C_i = \frac{2E_i}{k_i(k_i-1)},$$

where $E_i$ is the number of edges that actually exist between these $k_i$ neighbors of node $P_i$.

### 5.2. Global properties

The clustering coefficient of the whole network is the average of all individual clustering coefficients. Another important descriptor of the network as a whole is the *degree distribution* $P(k)$, defined as the proportion of nodes having degree $k$. The average degree $\langle k \rangle$ may be derived from it. The *path length* between any two nodes is defined as the minimal number of hops separating them. The diameter $L$ of the network is the maximal path length present in the network. Finally, the density of links $\rho$ is the ratio between the actual number of links present in the network and the number of links in a fully connected network [i.e., $N(N-1)/2$].

### 5.3. Random networks

In this work we need to compare the networks observed with random networks having the same number of nodes and links. There are several ways to obtain a random network with a specific number of nodes and links. The standard random networks introduced by Erdös and Rényi (Erdös 1959) simply distribute randomly the links between the nodes, keeping the number of nodes and links present in the original network which significance is to be tested for. However, this algorithm produces its own degree distribution, introducing a bias in the numerous cases where the degree distribution is not normal. To avoid this effect, the networks are usually randomized while keeping the degree distribution observed in the original network. In particular, starting from the original network, we picked two links and permuted the end nodes as described in reference (Maslov & Sneppen 2002). By repeating this procedure, we obtained uncorrelated random networks with the original degree distribution that allow testing for the significance of the original parameters.

## 6. Results and discussion

### 6.1. Genetic diversity spectrum

The meadows sampled differed greatly in clonal diversity, ranging from high monoclonal dominance (e.g., Es Castell, Spain, $R$=0.10) to highly diverse (e.g., Calabardina, Spain, $R$=0.88; Table 2, which is published as Supplementary information). The genetic distance between pairs of individuals within any population ranged from $D$=0 for clonal mates, to $D$=30 for the most genetically divergent individuals present in any population. The distribution of $D$ within any population is represented as a frequency distribution of all pairwise values, which we refer to as the *Genetic Diversity Spectrum* (GDS) of each population. The GDS is analogous to the frequency distribution of pairwise differences used on some clonal organism to detail the influence of clonality, as well as possible somatic mutations or scoring errors (Douhovnikoff & Dodd 2003; Meirmans & Van Tienderen 2004;

Van der Hulst et al. 2003). Nevertheless, we propose here to extend its interpretation beyond that particular application, using simulations to screen for the influence of some of the evolutionary forces that can contribute to shape the pattern of genetic diversity at the intra-population scale. In particular, we examine the importance of the mating system (outcrossing, selfing), which influences the way alleles are transmitted from one generation to the next, thereby playing a central role in the changing of allele frequencies across generations.

The GDS of the populations studied showed a range of shapes across populations (Fig. 1 and Fig. 5, which is published as supporting information). Three of the populations sampled (namely, Es Castell, Cala Fornells and Es Port) showed spectra indicative of a strongly clonal composition, characterized by a large spike at zero distance corresponding to the null distance between ramets pertaining to the same genetic individual, and discrete peaks located at characteristic genetic distances between the few distinct clones present in the population (Fig. 1a). Most of the populations sampled, however, are characterized by a broad, bimodal GDS (e.g., Fig. 1b), with a smaller (and broader) mode at zero distance, indicating the existence of nearly identical individuals forming clones, and a mode at higher distances within a broad, skewed, bell-shaped distribution.

The common characteristic features of the GDS from the different *P. oceanica* populations were highlighted by producing a mean GDS, obtained by averaging the GDS across all sampled populations. The resulting normalized histogram, which we call <GDS> and show in Fig. 2, is strongly bimodal, showing a large peak at 0 distance ($\alpha$ peak), suggesting that clonal reproduction constitutes one of the main factors influencing the intra-population genetic structure. The $\alpha$ peak is followed, at small genetic distances, by a depression, indicating that low genetic distances between 0.57 and 1,71 (corresponding to 4 to 12 nucleotides (nt) in case of genotypes composed by 7 loci) are uncommon. A broad peak ($\beta$ peak) at a modal pairwise genetic distance of 4.3 (approximately 30 nt) represents the most commonly observed genetic distance between genetically dissimilar (i.e., non-clonal) units sampled within populations. Above the $\beta$ peak distance, the frequency of distances between individuals declines exponentially (Fig. 2). Provided that enough polymorphic loci are used, which is assumed to be the case in our study where markers have been previously selected to that aim in a pilot work (Arnaud-Haond et al. 2005) the process responsible for generating genetic distances of 0 among individuals can be mostly assigned to clonal reproduction. In contrast, the processes generating specific classes of greater genetic distances are less apparent. However, this knowledge is essential for understanding, from a biological and mechanistic point of view, the implications of the observed ⟨GDS⟩ on the prevalence of various mechanisms that generate genetic diversity and structure within the population. The simulations performed allowed us to explore the range of genetic distances between parents and offspring, depending on the reproductive mode. The mean, across the populations examined, simulated genetic distance (±SE) generated by selfing and outcrossing was 1.97±0.16 and 3.43±0.17, respectively (Fig. 2). The characteristic genetic distance generated by simulated outcrossing ($\delta_{oc}= \delta^{(1)}_{oc}$) is close to the modal $\beta$ peak of the ⟨GDS⟩ (Fig. 2), suggesting that outcrossing is the main mechanism generating genetic diversity within the populations of this species. In contrast, the characteristic genetic distance generated by selfing ($\delta_s$) is located at the edge of the depression between the $\alpha$ and $\beta$ peaks in the ⟨GDS⟩, implying a low contribution of selfing in generating genetic distances in the meadows, and therefore a limited rate of selfing compared to outcrossing and clonality.

Observation of the ⟨GDS⟩ also suggests that caution should be exercised when interpreting the valley between the $\alpha$ and $\beta$ peaks in the spectrum of genetic distances in terms of somatic mutations or scoring errors, as proposed for obligatory outcrossing species (Douhovnikoff & Dodd 2003; Van der Hulst et al. 2003), when dealing with possibly self-fertilizing species. A small genetic distance can also be generated by selfing, a possibility that should be considered along with the more likely explanation that these distances arise from somatic mutations or scoring errors. In the case of possible self-fertilizers, simulations may be useful in defining the range of distances that can be generated sexually and the threshold below which clonality may be assumed. After such simulations in the case of *P. oceanica*, the uncommon distances between 0.29 and 0.86 (corresponding to 2 to 6 nt for the case of 7 loci genotypes) are still unlikely (data not shown) under a mixed mating system, with the selfing rate not exceeding the proportion expected on the basis of clone size (in terms of the number of shoots). Since we are confident that the double-checking procedure applied to the first data set allowed the correction of most scoring errors, these small distances must be mostly generated by somatic mutations accumulated in the process of multiple clonal reproductive events. Indeed, *P. oceanica* clones are extremely long-lived, with clones dated to millennia (Hemminga & Duarte 2000), over which clones would have divided multiple times, hence providing opportunities for somatic mutations. Indeed, the frequency of individuals at distances between strictly clonal ($D=0$) and the minimum observed in between the α and β peaks ($D=1.43$) also declines sharply, as expected from the low probability of accumulated mutations.

Lastly, the mean genetic distances from ancestors to offspring located *n* generations apart, obtained from simulations, increase very slowly with *n*, reaching an asymptote after approximately eight generations (Fig. 3). Comparison between the largest distances obtained by simulations and those observed on the ⟨GDS⟩ shows that the end of the distribution tail is not likely to be accounted for by sexual reproduction within the population. These distances are not likely to be generated by the random rearrangement of alleles during outcrossing or selfing over generations, but rather by external factors that generate diversity. The most likely process is a very low rate of immigration from other populations, which can suddenly introduce individuals genetically very distinct from those present in the population. However, examination of the contribution of immigration requires GDS evaluation across the entire distribution range of the species, rather than population-specific analyses such as that presented here.

**6.2. Network representation of the GDS**

The discussion above indicates that the GDS is best conceptualized as the result of genetic exchanges among a network of individual genets. Network analysis may, thus, provide a step forward in topologically characterizing the genetic relationships between population constituents depicted in the GDS. A first step to construct such network is to define the threshold genetic distance ($D_{th}$) representing closely genetically-connected individuals, characterized by between-individual distances ≤$D_{th}$. Based on analysis of the GDS, we choose to represent $D_{th}$ by the one-generation outcrossing distance (i.e., $D_{th}=\delta_{oc}$), which approximately corresponds to the β peak in the GDS of each population and is also very closely related to the percolation point. The network resulting from the connection of individuals at distances ≤$D_{th}$ represents the links among individuals that are approximately *up to a generation apart*, on the understanding that the genetic distance is only an operational *proxy* for the kinship among the individuals. In this work, the nodes in our networks are different genetic individuals, or genets, but we comment that networks of ramets can also be constructed, with a topology straightforwardly related to the one considered here. In the following we show the results for networks of genets, since the structure of ramet networks was straightforwardly obtained from them.

The networks (examples in Fig. 4) for the *P. oceanica* populations differ greatly in shape and in properties (Table 1) across the meadows analyzed.

As for the shape, the highly clonal population appears as a simple diagram of separate families with two or more clones each (Fig. 4a). In contrast, the network corresponding to the more diverse population is readily characterized by greater connectivity, showing a number of closely connected groups (sub-families) linked together by connections to a small set of central individuals, which act as links connecting the different families (Fig. 4b,d,e). In addition, we can distinguish fragmented (Fig. 4c) from connected (Fig. 4f) networks.

Comparing the properties, the largest component, *S*, of each network contains most of the individuals of each meadow (Table 1). The average degree of network connectivity ⟨*k*⟩ also differs greatly among populations (from 1.20 to 8.74, Table 1), with an overall average connectivity degree of 5.11, indicating that each individual is connected to, on average, five others. To indicate the significance of these numbers, we note that from the data in Table 1 the quantity ⟨*k*⟩/(*G*–1), which is the average proportion of the genet population connected to a given individual, ranges between 0.13 and 0.55. This implies that each individual is separated by at most one average generation from between 13% and 55% of the individuals in the sample. Together with the average link density, this shows that a large number of links are already present in the networks at the threshold chosen. Indeed, the density of links in the network averages 27% (Table 1), indicating that two randomly selected individuals in a given population have on average 27% probability of being less than one generation apart.

The average genetic link density of *P. oceanica* individuals (0.27; Table 1) is much higher than observed in other complex networks analyzed in the literature (Albert & Barabasi 2002). Similarly, the clustering coefficient $C$ (0.73; Table 1) is larger than observed in most reported networks, and generally larger than the clustering expected if the networks were random (i.e., $C > C_r$, Table 1). This departure from a random network signals the abundance of highly clustered nodes at the level of closely related families. The average path length ($L$), representing the minimum number of steps (links representing reproductive events or somatic mutations) necessary to connect any two individual genets in the population, ranged from 1.00 to 3.44, averaging 1.88 across populations (Table 1). This average path length suggests that most genets have a high kinship, typically below that of cousins, and is comparable to that generated by a random network. Taken together, the presence of short path lengths not departing from values expected in a random network ($L \approx L_r$) and of higher than expected clustering coefficient ($C > C_r$) indicate that the networks of genetic relationships in *P. oceanica* populations have the characteristics of a "small world" (Watts & Strogatz 1998). Small-world networks, as described extensively in the social sciences, characterize complex systems in which every node can be reached from every other using a small number of intermediate steps. This is indicative of a high degree of genetic substructure within populations of *P. oceanica*.

There is an interesting parallel between the sub-structure revealed by the networks shape and parameters, and the occurrence of spatial autocorrelation (Arnaud-Haond et al. in press) in some meadows (Cala Jonquet, Acqua Azzura 5; Fig. 4). Indeed, autocorrelation is used to detect patterns of limited dispersion revealed by a significant relationship between genetic and geographic distance, but a significant pattern primarily implies a sub-structure of the population in clusters of closely related individuals. This is very clearly illustrated by the shape of the networks, particularly in meadows where a strong pattern was detected, such as Cala Jonquet (Fig. 4b), where two subfamilies of five and nine highly interconnected individuals are linked together by a tiny path of three links and two intermediate nodes/individuals. However, failure to detect significant spatial autocorrelation does not imply the absence of sub-families in the population, but only a lack of relationship between this genetic structure and geographic distance, or else low statistical power. Exploration of the network of individuals can therefore reveal the existence of a sub-structure of the meadows in various families, if it exists, even when no pattern of spatial autocorrelation can be detected due to a lack of relationship between genetic and geographic distances. This is what is observed in most populations (Table 1) where a small world topology was detected, and the networks typically illustrate structures composed by highly connected sub-families (or clusters) inter-connected by few central nodes.

## 7. Conclusions

The results presented here illustrate a novel approach, based on analysis of the spectrum of genetic diversity, to examine the population genetic structure of clonal organisms and for the depiction of inter-individual genetic distances by a network. As underlined by Dyer and Nason (Dyer & Nason 2004), there are two fundamental distinctions between the classical genetic summary statistics, which involve decomposing variance, and the use of graphs. In the latter case, we do not impose pre-defined hierarchical models that constrain the range of temporal scales and evolutionary processes that can be accurately screened, but rather take advantage of all the information contained in the data set to let the data define their own topology and eventually offer a visual illustration. Here, we went one step further than graphical illustration by detailing the network properties using statistical tools specific to network analyses to extract key information on the hierarchical genetic structure in the population studied. This approach can be extended to explore the genetic structure of virtually any populations beyond the specific case of a marine clonal plant examined here. In doing so, many new elements have been introduced, such as a parsimony metric of distance among individual diploid organisms, the basis for the construction of the spectrum of genetic diversity. We also used simple simulations to explore the partition of the contribution of different processes to the genetic diversity contained in the spectrum, and topological representations of networks of genetic relationships derived from the spectrum of genetic diversity to formally explore the properties of the resulting network. Each of these novel approaches is rooted in earlier developments in different fields, such as computational population genetics (Meirmans & Van Tienderen 2004), population genetics (Bowcock et al. 1994; Dyer & Nason 2004), and network analysis developed in the realm of complex-systems theory (Albert & Barabasi 2002; Maslov & Sneppen 2002; Watts & Strogatz 1998), but are brought together here to provide a synthetic parsimony analysis of population genetic structures.

This interdisciplinary effort has allowed us to derive key features of the population genetics of the clonal species studied, such as the low contribution of somatic mutations and selfing to genetic diversity, and the inference, derived from examination of the spectrum of genetic diversity and subsequent network analysis, that *P. oceanica* populations show a rather high kinship level and low immigration of propagules produced in other populations. The network illustration allowed us to underline the high degree of sub-structure in some meadows, clearly composed of several families. The analysis of network properties allowed us to describe *P. oceanica* populations as following a typical "small-world" network shape, a feature already widely described in complex systems such as the World Wide Web and social networks (26), characterized by small diameters. A closer inspection to networks topology reveal that most of the meadows are composed by separated subgroups (families) interconnected through few "central" nodes.

The analysis of multiple populations has allowed elucidation of the considerable variability in the spectra of genetic diversity among populations, while identifying characteristic features in the spectrum. The approach demonstrated here can be extended to a wide range of organisms to explore genetic structure for a number of purposes. For instance, the topology of the genetic network for pathogens may help reveal important properties such as the clustering of strains, particularly when unusual transmission of genetic material such as lateral transfer are suspected to occur, and elucidate the evolutionary processes that shaped different lineages. The results presented here reveal the spectral and network analysis of genetic diversity as a promising tool to ascertain the genetic structure of populations and the role of different processes in shaping it.

**Acknowledgements**

This research was funded by a project of the BBVA Foundation (Spain), by project NETWORK (POCI/MAR/57342/2004) of the Portuguese Science Foundation (FCT) and CONOCE2 (fis2004-00953) of the Spanish MEC. S.A.H. was supported by a postdoctoral fellowship from FCT and the European Social Fund and A.F.R. by a post-doctoral fellowship of the Spanish Ministry of Education and Science. We also thank MARBEF European network and CORONA for fruitful group discussions.

**Figure Captions**

**Fig 1.** The Genetic Diversity Spectrum (GDS) for two representative populations (a) Es Castell, which is a strictly clonal population ($R$=0.1), and (b) Aqua Azzura 5, which has high clonal diversity ($R$=0.72).

**Fig 2.** The Genetic Diversity Spectrum averaged across sampled populations (⟨GDS⟩). The error bars indicate the SE for each bin. The square points (and corresponding error bars) were obtained from numerical simulations (see Methods) aimed at identifying mean (±SE) genetic distances generated by different biological processes: $\delta_c \equiv 0$ (clonal reproduction), $\delta_m$=0.86 (somatic mutations), $\delta_s$=1.89±0.11 (selfing, sexual reproduction between genetically identical individuals), $\delta_{oc}$ (outcrossing, sexual reproduction between genetically different individuals): $\delta_{oc}^{(1)}$=3.39±0.17, $\delta_{oc}^{(2)}$=4.24±0.23, $\delta_{oc}^{(4)}$=4.86±0.28, $\delta_{oc}^{(8)}$=5.05±0.29 and $\delta_{oc}^{(12)}$=5.1±0.29. The upper index indicates the number of generations apart for which the distance has been measured (1, 4, 8 and 12 generations). In the insert we show the same distribution on a log-linear scale. The straight line is a guide for the eye to highlight the exponential decay of the tail.

**Fig 3.** The cumulative distribution of genetic distances, obtained as the integral of the distribution shown in Fig. 2. We indicate the fraction of between-individual distances up to the values associated with different biological processes: mutation, selfing, and outcrossing of 1, 2, 4 and 8–12 generations. Insert: the whole range of genetic distances.

**Fig 4.** Network of genets for (a) Es Castell (Cabrera, Balearic Islands), (b) Cala Jonquet (Iberian Peninsula), (c) Rodalquilar (Iberian Peninsula), (d) Aqua Azzura5 (Sicily), (e) Roquetas (Iberian Peninsula) and (f)Playa Cavallets (Ibiza) after elimination of links representing genetic distances above the threshold $D_{th}=\delta_{oc}^{(1)}$. The value of $\delta_{oc}^{(1)}$ was obtained by means of numerical simulations and corresponds to the distance generated, on average, by outcrossing across one generation in the population. The node size is proportional to the number of identical constituent ramets.

**Fig 5a.** Supplementary material. We show the Genetic Diversity Spectrum for each sampled meadow: (1) Aqua Azzura 3, (2) Aqua Azzura 5, (3) Addaia, (4) Amathous 3, (5) Amathous 5, (6) Agios Nicolaos, (7) Calabardina, (8) Cala Giverola, (9) Cala Jonquet, (10) Campomanes, (11) Carboneras, (12) El Arenal, (13) Es Castell, (14) Es Pujols, (15) Es Calo de s'Oli, (16) Ses Illetes, (17) Cala Fornells, (18) La Fossa Calpe, (19) Las Rotes, and (20) Los Genoveces.

**Fig 5b.** Supplementary material (following); (21) Magaluf, (22) Malta, (23) Marzamemi, (24) Es Port, (25) Paphos, (26) Playa Cavallets, (27) Port Lligat, (28) Porto Colom, (29) Punta Fanals, (30) Rodalquilar, (31) Roquetas, (32) Cala Santa Maria 13, (33) Cala Santa Maria 7, (34) Cala Torreta, (35) Torre de la Sal, and (36) Tunis.

**Short title for page headings:**

"Genetic diversity and networks of clonal organisms"

|  | G | S | C | L | Cr | 90% CI (Cr) | Lr | 90% CI (Lr) | $\langle k \rangle$ | $\rho$ | Sp |
|---|---|---|---|---|---|---|---|---|---|---|---|
| Acqua Azzura 3 | 31 | 28 | **0.80** | 3.13 | 0.22 | [0.15,0.29] | 2.1 | [2.12,2.22] | 4.71 | 0.16 | **0.02** |
| Acqua Azzura 5 | 29 | 25 | **0.78** | 2.43 | 0.25 | [0.19,0.31] | 2.0 | [1.95,2.06] | 4.76 | 0.17 | **0.01** |
| Addaia | 25 | 17 | **0.72** | 1.64 | 0.41 | [0.34,0.48] | 1.9 | [1.87,2.01] | 5.60 | 0.23 | **0.02** |
| Agios | 28 | 18 | **0.81** | 2.41 | 0.32 | [0.23,0.41] | 1.8 | [1.81,1.94] | 5.50 | 0.20 | **0.06** |
| Amathous 3 | 18 | 11 | **0.77** | 1.45 | 0.27 | [0.19,0.36] | 2.1 | [2.03,2.33] | 4.67 | 0.27 | 0.01[NS] |
| Amathous 5 | 25 | 24 | **0.73** | 3.14 | 0.25 | [0.19,0.32] | 2.1 | [2.07,2.19] | 4.56 | 0.19 | 0.00[NS] |
| Calabardina | 40 | 40 | **0.62** | 3.44 | 0.18 | [0.13,0.24] | 2.2 | [2.17,2.31] | 5.90 | 0.15 | 0.00[NS] |
| Cala Giverola | 17 | 10 | **0.84** | 1.22 | 0.72 | [0.67,0.77] | 1.6 | [1.55,1.77] | 4.35 | 0.27 | 0.01[NS] |
| Cala Jonquet | 20 | 18 | **0.79** | 2.95 | 0.36 | [0.27,0.46] | 1.9 | [1.84,1.97] | 4.60 | 0.24 | **0.07** |
| Campomanes | 22 | 12 | **0.66** | 1.77 | 0.26 | [0.14,0.40] | 2.1 | [2.01,2.28] | 4.00 | 0.19 | -0.01[NS] |
| Carboneras | 16 | 16 | 0.78 | 1.75 | 0.73 | [0.68,0.78] | 1.5 | [1.48,1.55] | 8.50 | 0.57 | 0.03[NS] |
| El Arenal | 32 | 27 | **0.76** | 2.74 | 0.28 | [0.21,0.36] | 2.1 | [1.99,2.27] | 5.25 | 0.17 | **0.02** |
| Es Castell | 05 | 03 | 0.00 | 1.33 | 0.00 | - | 2.1 | - | 1.20 | 0.30 | **0.12** |
| Es Pujols | 27 | 24 | **0.74** | 2.76 | 0.57 | [0.50,0.64] | 1.9 | [1.84,2.04] | 5.70 | 0.22 | 0.01[NS] |
| Es Calo de s'Oli | 15 | 07 | **0.73** | 1.14 | 0.47 | [0.36,0.60] | 1.9 | [1.74,2.11] | 2.80 | 0.20 | 0.00[NS] |
| Ses Illetes | 21 | 20 | **0.78** | 2.62 | 0.36 | [0.28,0.46] | 1.8 | [1.78,1.88] | 5.52 | 0.28 | - |
| Cala Fornells | 05 | 03 | 1.00 | 1.00 | 1.00 | - | 1.0 | - | 1.20 | 0.30 | -0.06[NS] |
| La Fossa | 31 | 29 | **0.63** | 2.76 | 0.24 | [0.19,0.29] | 2.0 | [1.96,2.16] | 5.55 | 0.18 | 0.00[NS] |
| Las Rotes | 34 | 21 | **0.75** | 1.98 | 0.20 | [0.13,0.27] | 2.2 | [2.20,2.32] | 4.65 | 0.14 | **0.04** |
| Los Genoveces | 14 | 13 | 0.86 | 1.45 | 0.84 | [0.82,0.87] | 1.3 | [1.34,1.43] | 7.71 | 0.59 | 0.03[NS] |
| Magaluf | 26 | 18 | **0.64** | 1.88 | 0.39 | [0.29,0.50] | 2.1 | [2.03,2.24] | 4.46 | 0.18 | 0.02[NS] |
| Malta | 29 | 24 | **0.67** | 2.19 | 0.33 | [0.26,0.40] | 1.9 | [1.85,1.97] | 5.24 | 0.19 | - |
| Marzamemi | 31 | 28 | **0.63** | 2.48 | 0.39 | [0.31,0.48] | 2.0 | [1.95,2.15] | 5.61 | 0.19 | **0.01** |
| Es Port | 05 | 05 | 0.90 | 1.10 | 0.90 | - | 1.1 | - | 3.60 | 0.90 | -0.01[NS] |
| Paphos | 26 | 17 | **0.74** | 1.65 | 0.41 | [0.34,0.49] | 1.9 | [1.91,2.04] | 6.54 | 0.26 | 0.01[NS] |
| Playa Cavallets | 28 | 24 | **0.63** | 2.45 | 0.32 | [0.22,0.42] | 1.9 | [1.91,2.01] | 4.86 | 0.18 | 0.00[NS] |
| Port Lligat | 12 | 07 | **0.95** | 1.10 | 0.39 | [0.27,0.52] | 1.5 | [1.55,1.64] | 4.17 | 0.38 | 0.01[NS] |
| Porto Colom | 21 | 16 | **0.83** | 1.42 | 0.72 | [0.66,0.78] | 1.5 | [1.38,1.65] | 7.62 | 0.38 | **0.06** |
| Punta Fanals | 26 | 26 | **0.70** | 2.06 | 0.47 | [0.41,0.53] | 1.8 | [1.80,1.93] | 7.54 | 0.30 | -0.01[NS] |
| Rodalquilar | 27 | 14 | **0.84** | 1.32 | 0.37 | [0.32,0.42] | 1.7 | [1.73,1.81] | 8.22 | 0.32 | 0.00[NS] |
| Roquetas | 35 | 34 | **0.79** | 2.47 | 0.33 | [0.29,0.37] | 1.8 | [1.76,1.89] | 8.74 | 0.26 | 0.01[NS] |
| C.Sta.Maria | 20 | 19 | **0.73** | 2.37 | 0.42 | [0.33,0.50] | 1.7 | [1.74,1.84] | 5.50 | 0.29 | 0.01[NS] |
| C.Sta.Maria 7 | 22 | 16 | **0.66** | 1.74 | 0.57 | [0.48,0.65] | 1.9 | [1.79,2.03] | 4.91 | 0.23 | 0.01[NS] |
| Cala Torreta | 21 | 10 | **0.75** | 1.84 | 0.13 | [0.02,0.24] | 2.4 | [2.28,2.61] | 3.33 | 0.17 | **0.06** |
| Torre de la Sal | 15 | 13 | **0.71** | 1.64 | 0.58 | [0.52,0.64] | 1.5 | [1.37,1.64] | 4.93 | 0.35 | 0.03[NS] |
| Tunis | 34 | 30 | **0.77** | 2.80 | 0.27 | [0.20,0.34] | 2.1 | [2.10,2.24] | 5.41 | 0.16 | - |
| Xilxes | 12 | 05 | 0.47 | 1.80 | 0.29 | [0.03,0.56] | 1.9 | [1.24,2.71] | 1.50 | 0.14 | - |
| $\langle x \rangle$ | 22.8 | 18.16 | 0.73 | 2.04 | 0.41 | - | 1.8 | - | 5.11 | 0.27 | - |
| $\sigma$ | 8.54 | 8.82 | 0.16 | 0.65 | 0.22 | - | 0.3 | - | 1.77 | 0.15 | - |
| min($x$) | 5.00 | 3.00 | 0.00 | 1.00 | 1.00 | - | 2.4 | - | 1.20 | 0.14 | - |
| max($x$) | 40.0 | 40.00 | 1.00 | 3.44 | 0.00 | - | 1.0 | - | 8.74 | 0.90 | - |

**Table 1.** Summary of properties measured for population genetic networks of genets (non-similar ramets) at $D_{th}=\delta_{oc}$. $G$ is the number of genets present in the meadow, $S$ stands for the size of major connected components, $C$ for the clustering, $L$ for the diameter, Cr, Lr, 90% CI (Cr) and 90% CI (Lr) for the average clustering and diameter and their 90% confidence interval, after random rewiring, $\langle k \rangle$ for the mean degree of connectivity, $\rho$ for the link density and $Sp$ for the spatial autocorrelation $Sp$ statistics. For $C$, $L$ and $Sp$, bold

values indicate significance as departure from the simulated random distributions ($p<0.05$). In the last four lines of the table we statistically characterize each column: <x> stands for the mean and σ for the standard deviation. In the last two lines we show the minimum and maximum values.

| | Locality | Lat. | Long. | SUs | *R* | *Sp* |
|---|---|---|---|---|---|---|
| SPAIN (Iberic Peninsula) | | | | | | |
| | Roquetas | 36° 43.26'N | 2° 37.09'W | 50 | 0.69 | 0.01[NS] |
| | Rodalquilar | 36° 51.21'N | 2° 00.53W | 50 | 0.53 | 0.00[NS] |
| | Los Genoveces | 36° 44.40'N | 2° 07.02W | 39 | 0.34 | 0.03[NS] |
| | Carboneras | 36° 69.61'N | 1° 53.20'W | 48 | 0.32 | 0.03[NS] |
| | Calabardina | 37° 26.00'N | 1° 30.00'W | 48 | 0.88 | 0.00[NS] |
| | Campomanes | 38° 37.54' N | 0° 0.57'E | 31 | 0.7 | - |
| | Torre de la Sal | 40° 8.13' N | 0° 10.72'E | 39 | 0.5 | 0.03[NS] |
| | El Arenal | 38° 38.37' N | 0° 3.06'E | 39 | 0.86 | 0.02* |
| | La Fossa Calpe | 38° 33.59'N | 0° 4.56'E | 40 | 0.77 | 0.00[NS] |
| | Xilxes | 39° 45.13' N | 0° 8.07'E | 32 | 0.35 | - |
| | Las Rotes | 38° 50.03' N | 0° 8.56'E | 50 | 0.73 | 0.04* |
| | Punta Fanals | 41° 41.58'N | 2° 50.56'E | 38 | 0.68 | - |
| | Cala Giverola | 41° 44.15'N | 2° 57.37'E | 38 | 0.43 | 0.01[NS] |
| | Cala Jonquet | 42° 18.19'N | 3° 17.36'E | 39 | 0.50 | 0.07* |
| | Port Lligat | 42° 17.61'N | 3° 17.58'E | 40 | 0.28 | 0.01[NS] |
| SPAIN (Balearic Islands) | | | | | | |
| Ibiza | Playa Cavallets | 38° 50.99'N | 1° 24.25'E | 38 | 0.73 | 0.00[NS] |
| Formentera | Es Calo de s'Oli | 38° 43.49'N | 1° 24.16'E | 40 | 0.36 | 0.00[NS] |
| | Cala Torreta | 38° 47.45'N | 1° 25.18'E | 40 | 0.51 | 0.06* |
| | Ses Illetes | 38° 45.37'N | 1° 25.83'E | 36 | 0.60 | - |
| | Es Pujols | 38° 43.74'N | 1° 27.27'E | 40 | 0.67 | 0.01[NS] |
| Cabrera | Es Castell | 39° 9.16'N | 2° 55.83'E | 40 | 0.10 | 0.12* |
| | Es Port | 39° 8.81'N | 2° 55.86'E | 40 | 0.10 | - |
| | C. Sta. María 13 m | 39° 9.07'N | 2° 56.92' | 35 | 0.56 | 0.01[NS] |
| | C. Sta. María 7 m | 39° 9.00'N | 2° 56.96'E | 40 | 0.54 | 0.01[NS] |
| Mallorca | Magaluf | 39° 30.25'N | 2° 32.59'E | 38 | 0.68 | 0.02[NS] |
| | Porto Colom | 39° 25.05'N | 3° 16.18'E | 41 | 0.50 | 0.06* |
| Menorca | Cala Fornells | 40° 03.39'N | 4° 08.26'E | 40 | 0.10 | - |
| | Addaia | 40° 00.97'N | 4° 12.42'E | 37 | 0.67 | 0.02* |
| TUNISIA | Tunis | 36° 46.00'N | 10° 19.00'E | 40 | 0.85 | - |
| MALTA | Malta | 35° 51.00'N | 14° 35.00'E | 39 | 0.74 | - |
| ITALY (Sicily) | Acqua Azzurra 3 | 36° 42.71'N | 15° 8.44'E | 40 | 0.77 | 0.02* |
| | Acqua Azzurra 5 | 36º 43.31'N | 15º8.48'E | 40 | 0.72 | 0.01* |
| | Marzamemi | 36° 43.29'N | 15° 0.49'E | 38 | 0.81 | 0.01* |
| GREECE | Agios Nicolaos | 37° 42.97'N | 23° 55.62'E | 40 | 0.69 | 0.06* |
| CYPRUS | Amathous 3 | 34° 41.96'N | 33° 12'00'E | 40 | 0.44 | 0.01[NS] |
| | Amathous 5 | 34° 42.02'N | 33° 12.99' | 40 | 0.62 | 0.00[NS] |
| | Paphos | 34° 43.54'N | 32° 26.23'E | 38 | 0.68 | 0.01[NS] |

**Table 2.** Supplementary material. Sampling details: country, locality, approximate GPS coordinates, number of sampling units collected (SUs), genotype diversity (*R*).

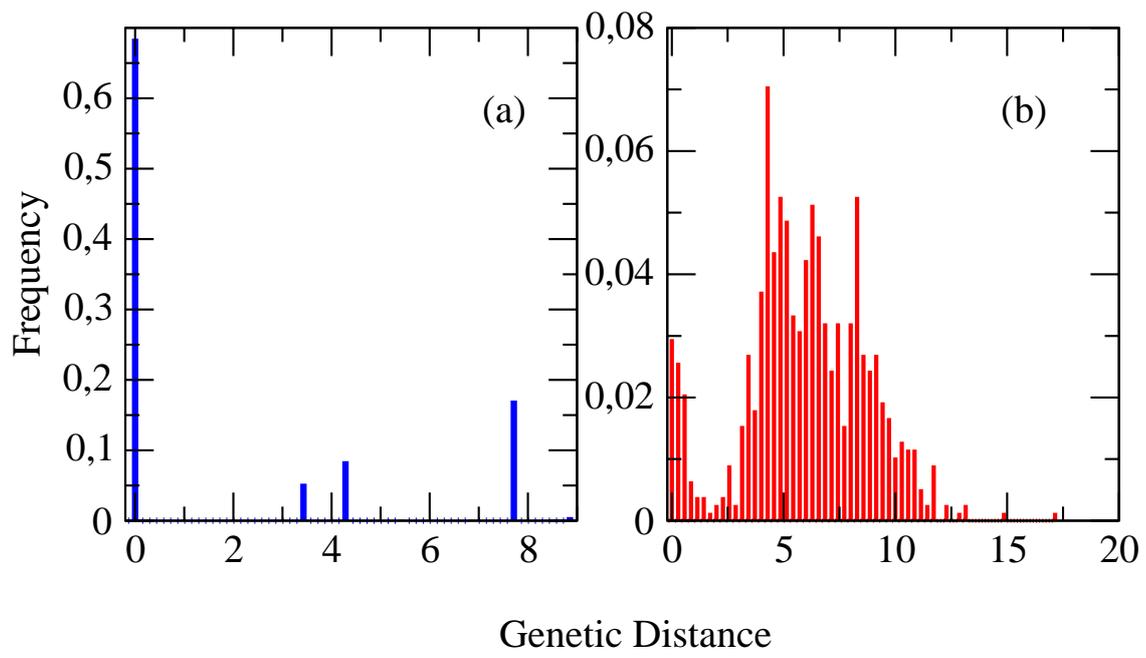

Figure 1.

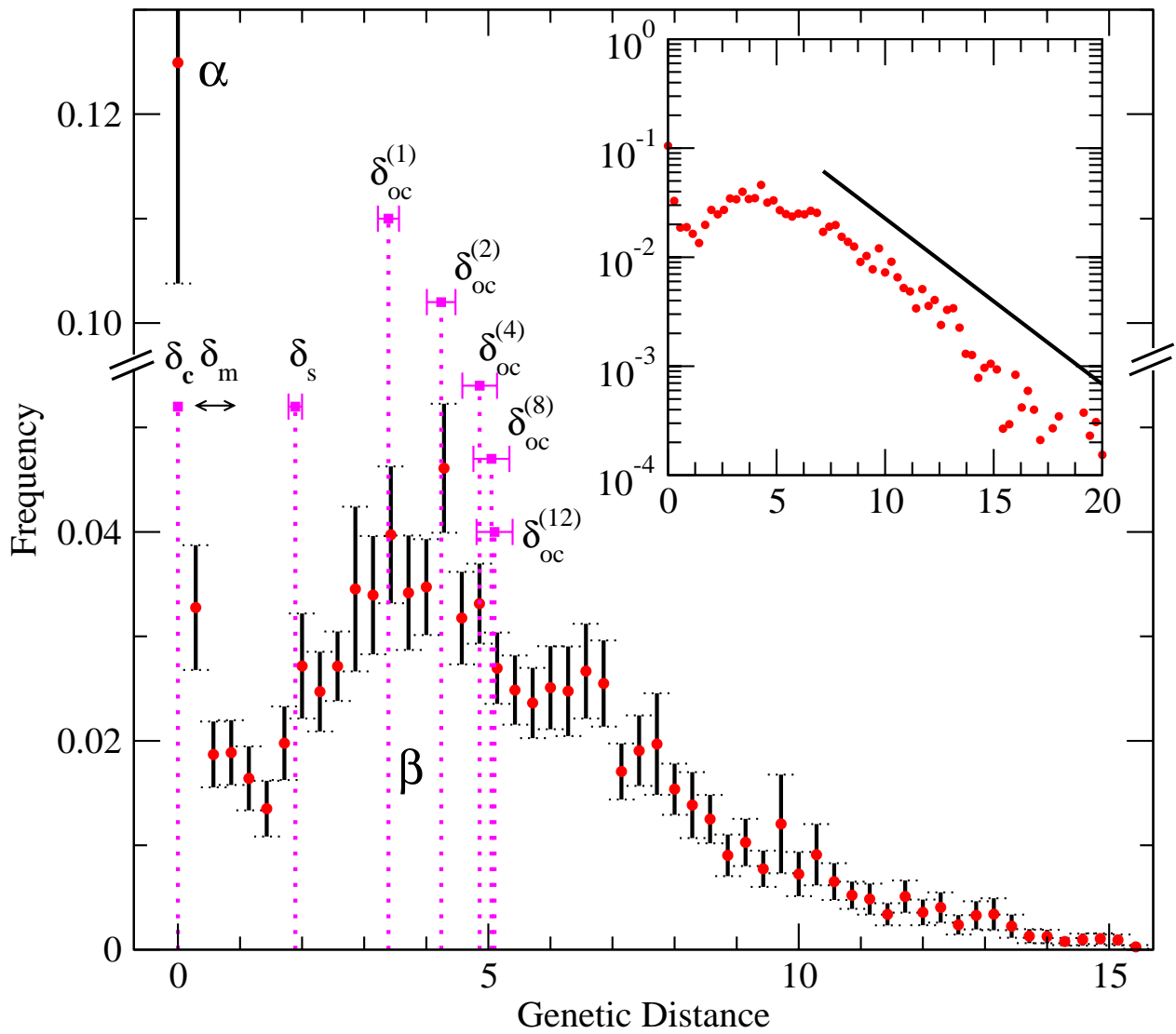

Figure 2.

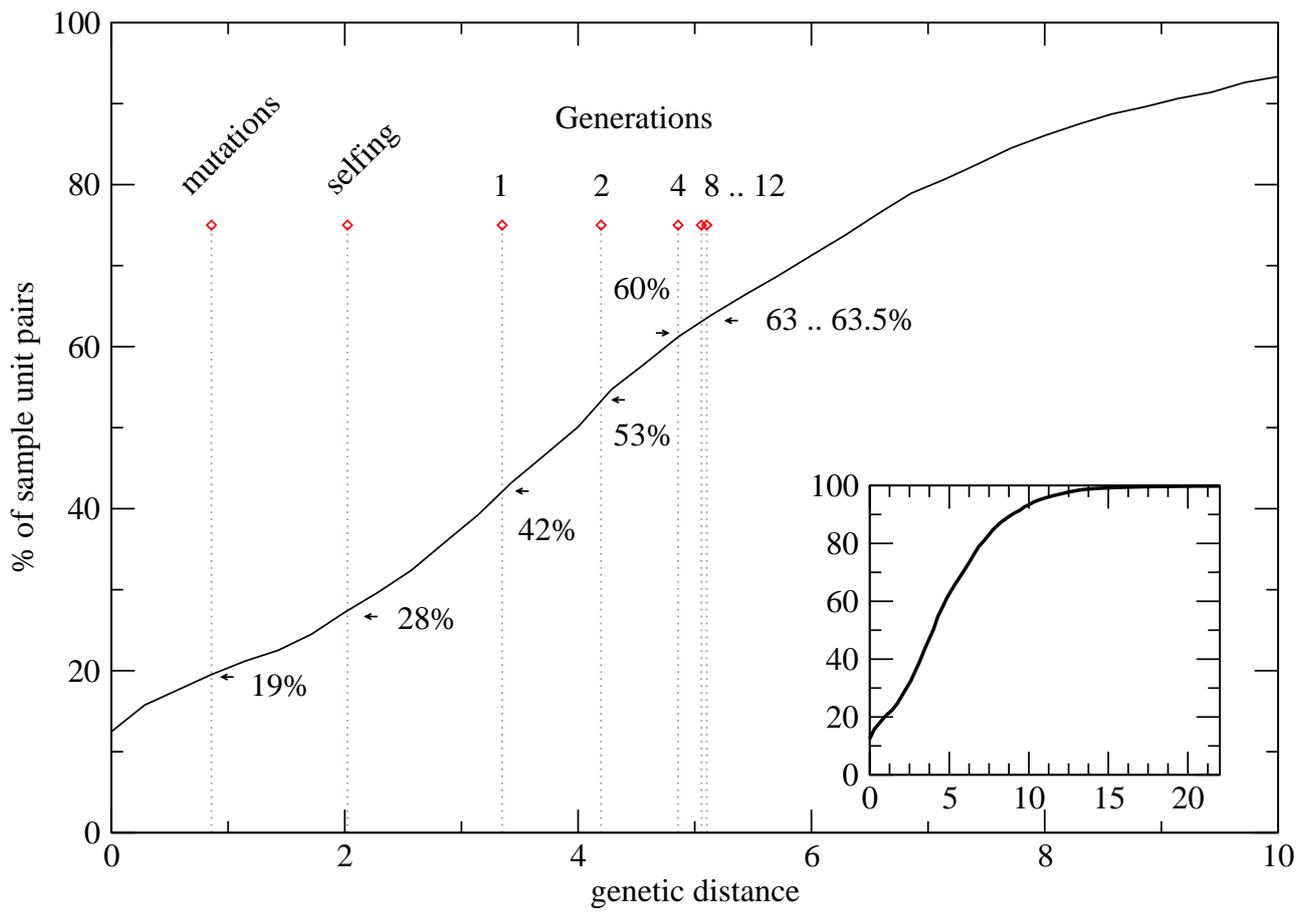

Figure 3.

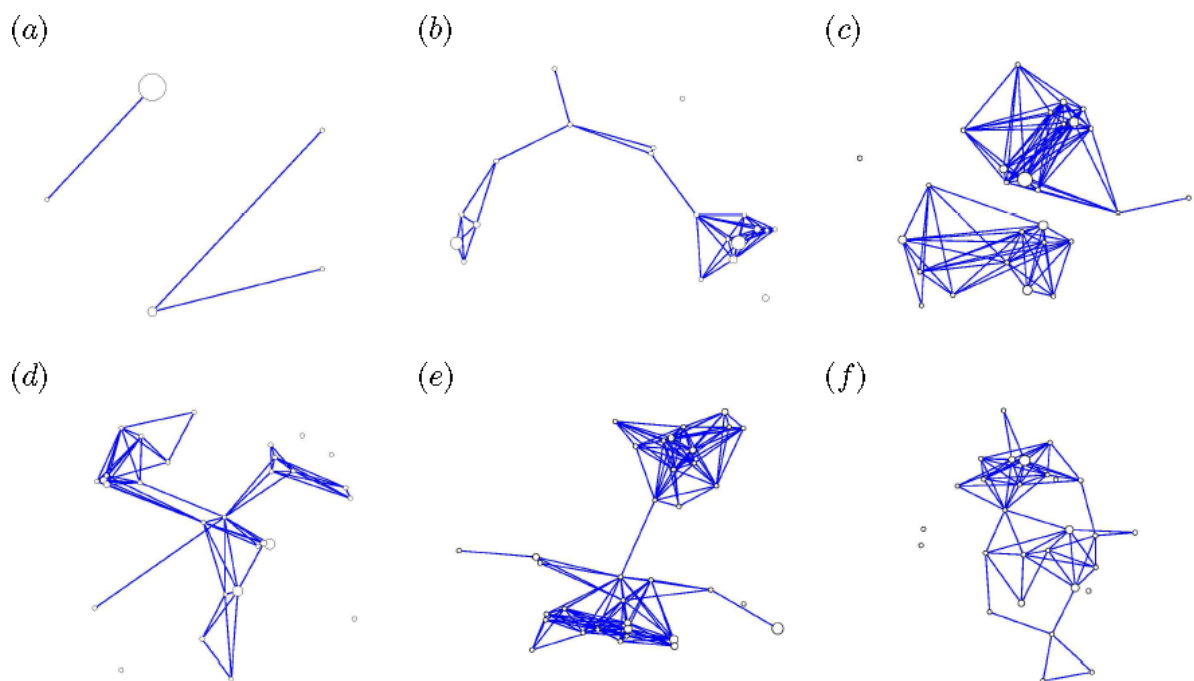

Figure 4.

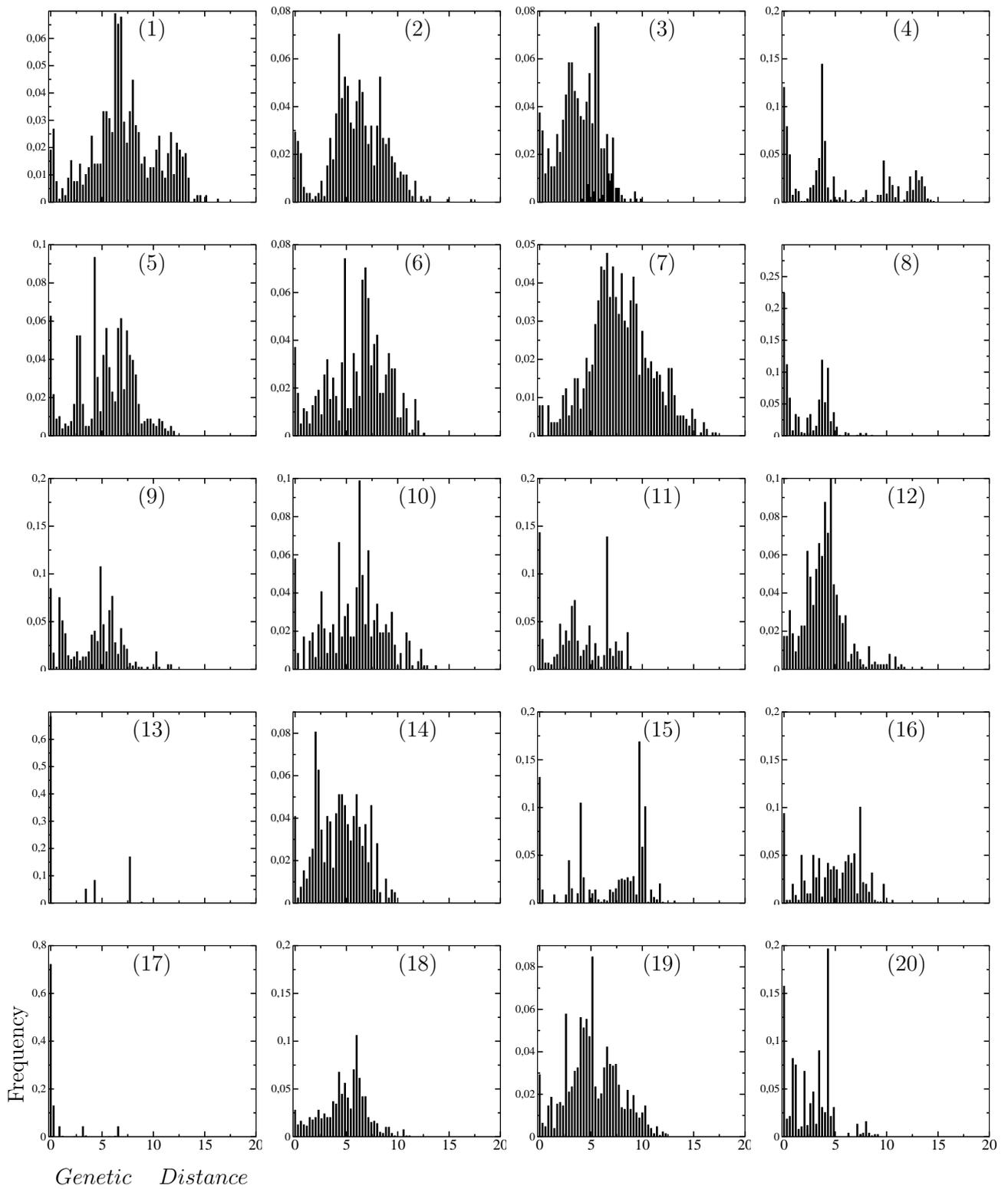

Figure 5 (a). Supplementary material.

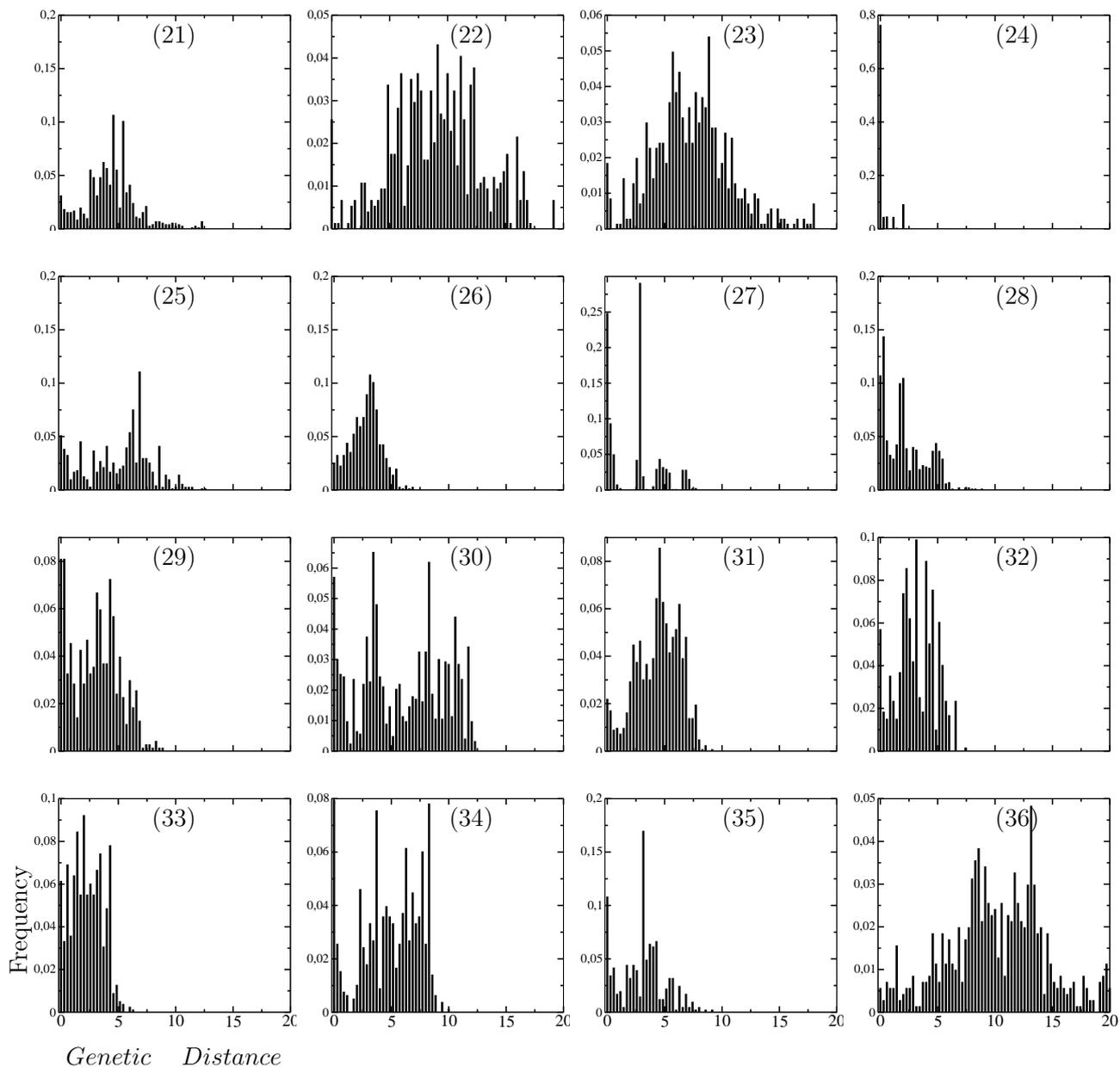

Figure 5 (b). Supplementary material.